\pdfoutput=1

\documentclass[11pt,twoside,a4paper,cmspaper,final,collab]{cms-tdr}
\def\svnVersion{80117:80133}\def\cmsCernNo{2011-086}\def\cmsCernDate{\today}\def\cmsMessage{Submitted to the Journal of High Energy Physics}

\begin{document}\cmsNoteHeader{FWD-10-011}

\hyphenation{had-ron-i-za-tion}
\hyphenation{cal-or-i-me-ter}
\hyphenation{de-vices}
\RCS$Revision: 80128 $
\RCS$HeadURL: svn+ssh://alverson@svn.cern.ch/reps/tdr2/papers/FWD-10-011/trunk/FWD-10-011.tex $
\RCS$Id: FWD-10-011.tex 80128 2011-10-02 17:51:00Z alverson $
\cmsNoteHeader{FWD-10-011} 
\title{Measurement of energy flow at large pseudorapidities in \Pp\Pp\ collisions at $\sqrt{s}$  = 0.9 and 7 TeV }

\date{\today}

\abstract{
The energy flow, $dE/d\eta$, is studied at large pseudorapidities
in proton-proton collisions at the LHC, for centre-of-mass energies of 0.9 and 7~TeV. The measurements are made in the pseudorapidity range $3.15 < |\eta| <  4.9$, for both minimum-bias events and events with at least two high-momentum jets, using the CMS detector. The data are compared to various \Pp\Pp~Monte Carlo event generators whose theoretical models and input parameter values are sensitive to the energy-flow measurements.  Inclusion of multiple-parton interactions in the Monte Carlo event generators is found to improve the description of the energy-flow measurements.
}

\hypersetup{%
pdfauthor={CMS Collaboration},%
pdftitle={Measurement of energy flow at large pseudorapidities in pp collisions at sqrt(s)  = 0.9 and 7 TeV},%
pdfsubject={CMS},%
pdfkeywords={CMS, physics, HF, energy flow}}

\maketitle 

\def\question#1{\footnote{\textbf{QUESTION: #1}}}
\def\answer#1{\footnote{\textbf{ANSWER: #1}}}
\newcommand{\todo}[1]{\vspace*{3mm}\par\noindent\textbf{\boldmath $\Longrightarrow$ #1}\vspace*{4mm}}
\newcommand{\DGLAP}{Gribov:1972ri,Lipatov:1974qm,Altarelli:1977zs,Dokshitzer:1977sg}
\newcommand{\BFKL}{Kuraev:1976ge,Kuraev:1977fs,Balitsky:1978ic}
\newcommand{\CCFM}{Ciafaloni:1987ur,Catani:1989yc,Catani:1989sg,Marchesini:1994wr}

\newcommand{\update}[0]{\textbf{"to be updated" }}

\section{Introduction}
\label{sec:Introduction}
Proton-proton collisions at the Large Hadron Collider (LHC) allow Quantum Chromodynamics (QCD) processes to be investigated through the measurement of the average energy per event (energy flow) in specific angular regions. Such a measurement is useful in examining the complex final states that result from a hadron-hadron interaction. Exploiting the large calorimeter coverage of the Compact Muon Solenoid (CMS) detector  allows the energy flow to be measured over a wider range than was accessible in previous analyses.

At the LHC, the accessible fraction $x$ of the proton momentum carried by partons can become very small. At small $x$, parton densities become large, leading to an increased probability for more than one partonic interaction. The final state resulting from
a proton-proton interaction
 can be described as the superposition of several contributions:
 the hard interaction,
 initial-state and final-state radiation, hadrons produced in additional multiple-parton interactions~\cite{Sjostrand:1987su}, and beam-beam remnants resulting from the hadronisation of the partons that did not participate in the hard scatter.
The underlying event (UE) is defined as everything except the hard interaction, i.e.,
multiple-parton interactions, beam-beam remnants,  and initial-state and final-state radiation. The UE plays an especially important role at high centre-of-mass energy.

Multiple-parton interactions  are not well understood theoretically
and a systematic description in QCD remains challenging \cite{Bartalini:2010su}. Phenomenological approaches to multi-parton dynamics rely  strongly
on parameterised models and tuned parameters (tunes) to describe data.
Measurements of the UE structure have been performed for central values of the pseudorapidity ($\eta$), where $\eta=-\ln\left[\tan\left(\theta/2\right)\right]$, with $\theta$ being the polar angle of the particles with respect to the beam axis ~\cite{Acosta:2004wqa,QCD-10-001,Chatrchyan:2011id}.
The energy flow has been measured previously in \Pp\Pap\ collisions~\cite{Albajar:1989an} at $\sqrt{s}=0.2 - 0.9$~TeV, as well as in \Pe\Pp\ collisions~\cite{Adloff:1999ws}. The extension of the measurements to large pseudorapidities and higher centre-of-mass energies is a challenge for the models since, in this region of phase space, parton showers (initial-state radiation), as well as multiple-parton interactions, are expected to play a significant role.

In this paper, the measurement of the energy flow in the pseudorapidity range $3.15 <|\eta|< 4.9$ is presented. The energy flow in an event is defined as $dE / d\eta  = \sum_i E_i / \Delta \eta $, where $\sum_i E_i$ is the summed energy of all  charged and neutral particles in the event
measured in bins of pseudorapidity, and $\Delta \eta$ is the bin width in $\eta$. Two different centre-of-mass energies, 0.9 and 7~TeV, were investigated for two different event classes: minimum-bias events and events with central high-transverse-momenta dijets. The latter are expected
to be more sensitive to perturbative QCD phenomena.

This paper is organised as follows: Section \ref{sec:Detector} describes the experimental apparatus, and the event selection is explained in Section \ref{sec:EventSelection}. The main features of the Monte Carlo (MC)  event generators used for the comparison with data are presented in Section \ref{sec:MC}. Sections~\ref{sec:Corrections} and~\ref{sec:Systematics} discuss the corrections and systematic uncertainties related to the energy-flow measurement. The results are presented in Section \ref{sec:Results}, and the final conclusions are summarised in Section~\ref{sec:Conclusions}.

\section{Experimental Apparatus}
\label{sec:Detector}

This section briefly summarises some features of the CMS apparatus relevant for the present measurement. A complete description of the CMS detector can be found elsewhere~\cite{JINST}. The CMS experiment uses a right-handed coordinate system, with the origin at the nominal interaction point (IP), the $x$ axis pointing towards the centre of the LHC ring, the $y$ axis pointing upwards, and the $z$ axis along the anticlockwise-beam direction. The azimuthal angle $\phi$ is measured with respect to the $x$ axis, in the $x-y$ plane, and the polar angle $\theta$ is defined with respect to the $z$ axis.

The central feature of CMS is a superconducting solenoid, of 6~m internal diameter, providing a magnetic field of 3.8~T. Within the field volume are the silicon pixel and strip tracker, the crystal electromagnetic calorimeter  and the brass/scintillator hadron calorimeter. The tracker measures charged particle trajectories in the pseudorapidity range $|\eta| <$ 2.5.

For triggering purposes, the CMS trigger system~\cite{Bayatian:2006zz,mbtrigger} was used, together with two elements of the CMS detector monitoring system, the beam scintillation counters (BSC)~\cite{Bell2008} and the beam Pick-up-timing for the experiments (BPTX) devices~\cite{bptx2}. A BSC detector is located on each side of the interaction region, at a distance of 10.86~m from the nominal crossing point,
covering the $|\eta|$ range from 3.23 to 4.65. Each BSC consists of a set of 16 scintillator tiles. They provide information on hits and coincidence signals with an average detection efficiency of 96.3\% for minimum-ionising particles and a time resolution of 3 ns, compared to a minimum inter-bunch spacing of 50 ns for collisions. Located around the beam pipe at distances of 175~m on either side of the IP, the two BPTX devices are designed to provide precise information on the structure and timing of the LHC beams, with a time resolution better than 0.2 ns.

For the present analysis, the energy was measured with the hadronic forward (HF) calorimeters, which cover the region  2.9$< |\eta|<$5.2. The front face of each calorimeter is located 11.2~m from the interaction point, at each end of CMS. The HF calorimeters consist of iron absorbers and embedded radiation-hard quartz fibres, providing a fast collection of the \v Cherenkov light. The collected light is detected using radiation-hard photomultiplier tubes (PMT). The quartz fibres are grouped into two categories, segmenting the HF calorimeters longitudinally: long fibres, which run over the full depth of the absorber (165 cm $\approx$ 10$\lambda_I$), and short fibres, which start at a depth of 22 cm from the front of the detector. Each set of fibres is read out separately. The different lengths of the fibre sets make it possible to distinguish between electromagnetic and hadronic showers.

Calorimeter cells are formed by grouping bundles of fibres. Clusters of these cells (e.g.,   3$\times$3 grouping) form a calorimeter tower. There are 13 towers in $\eta$, each with a size given by $\Delta\eta \approx$ 0.175, except for the lowest- and highest-$|\eta|$ towers with $\Delta\eta \approx$ 0.1 and $\Delta\eta\approx$ 0.3, respectively. The azimuthal segmentation of all towers is 10$^\circ$, except for the one at highest-$|\eta|$, which has $\Delta\phi=$20$^\circ$.

Corrections were applied to the data in order to account for geometrical non uniformities of the HF calorimeters, due to non sensitive areas. These are caused by mechanical structures of the 20$^\circ$ $\phi$-segment assembly, which are not included in the detector simulation. The relevant correction factors for the non uniformities range between 0.98 for the innermost (large $\eta$) pseudorapidity segment and 0.90 for the outermost (small $\eta$) segment, and were applied to the total energy deposited in each tower.

The calibration of the HF calorimeters is based on test-beam measurements with 100~GeV pions and electrons~\cite{Abdullin}. For the relative calibration, a $^{60}$Co radioactive source was used.
The energy-flow analysis was restricted to the pseudorapidity range 3.15$<|\eta|<$4.9. The lowest and highest $|\eta|$ towers were excluded from the measurement since  they were not properly described in the geometry used for the detector simulation. An additional ring of towers was removed from the analysis
because it is partially located in the shadow of the electromagnetic endcap calorimeter.

\section{Event Selection}
\label{sec:EventSelection}
The data were collected during the first data-taking period of CMS in 2010, corresponding to integrated luminosities of 239~$\mu $b$^{-1}$ and 206~$\mu$b$^{-1}$  for the  $\sqrt{s}$ =  0.9 and 7~TeV samples, respectively.
In less than 1\% of the events more than one interaction vertex was recorded (pile up).
The energy was measured with the HF calorimeters by summing all energy deposits in the HF towers above a threshold of 4~GeV. This threshold, determined from non collision events,
was chosen to suppress electronic noise.
In addition, events in which particles hit the photomultipliers and cause large signals in the HF calorimeter towers were removed from the analysis
by means of dedicated
algorithms. The rejection criteria are based on the topology of energy deposits and the pulse shape/timing of the signals in the HF calorimeters.

\subsection{Minimum-bias Events}
Minimum-bias events were selected by requiring a trigger signal from the two BPTX detectors indicating the presence of both beams crossing the interaction point, coincident with a signal in each of the BSC detectors. At least one reconstructed primary vertex was required, with a $z$ coordinate within 15 cm of the centre of the beam collision region. The vertex was reconstructed in a fit with a minimum of four associated tracks~\cite{Collaboration:2010xs}.
Beam-induced background events producing an anomalously large number of pixel hits were rejected by requiring that at least 25\% of the tracks in events with more than 10 tracks were of high quality.
Tracks were considered of high quality when several requirements were fulfilled,
such as the normalised $\chi^2$ of the fit, the compatibility with the primary vertex, and the number of hit layers. A detailed description of the high quality criteria can be found in Ref.~\cite{TRK-10-001}.

A large number of single-diffraction events was removed with the BSC trigger requirement.
The fraction of events from single- and double-diffractive dissociation remaining in the samples was estimated with a {\sc pythia}6 MC event sample: 5\% (3\%) for $\sqrt{s}$ = 0.9~TeV (7~TeV).
After applying all the selection criteria, 2.8 $\times$ 10$^6$ and 9.4 $\times$ 10$^6$ events remained in the $\sqrt{s}$ = 0.9~TeV and $\sqrt{s}$ = 7~TeV data samples, respectively.

\subsection{Dijet Events\label{qcdsample}}
The jets were reconstructed using the anti-$k_T$ jet clustering algorithm~\cite{Cacciari:2008gp}, with a distance parameter of 0.5.
Particle flow objects~\cite{particleflow}  that combine information from all the CMS sub-detectors were used as input.
The reconstructed jets required only small additional energy corrections derived from the {\sc pythia}6 MC generator and in-situ measurements with photon+jet and dijet events~\cite{JME-10-010}. The contribution of miss-reconstructed jets was kept at a negligible level by means of the procedures discussed in Refs.~\cite{JME-10-001,JME-10-003}.

The dijet event sample used in the analysis was a subset of the minimum-bias sample. For the event selection, the two highest-$p_{T}$ jets were required to satisfy the condition $|\Delta\phi(jet_{\textrm{1}},jet_{\textrm{2}})-\pi|<$ 1.0, where $\Delta\phi(jet_{\textrm{1}},jet_{\textrm{2}})$ is the difference in the azimuthal angles of the two jets, and to lie within the central region ($|\eta| < $  2.5). At $\sqrt{s}$  =  0.9~TeV ($\sqrt{s}$ = 7~TeV), the leading and sub-leading jets were both required to have $p_{T,jet} >$ 8~GeV ($p_{T,jet} > $ 20~GeV). The requirement on $\eta$ ensures that the jets are contained in the central region of CMS, outside the acceptance of the HF calorimeters. The requirement on the minimum transverse momentum of the jets was chosen to reconstruct the jets fully efficiently. The threshold
choices result in an approximately similar lower limit on the fractional momentum carried by the jets,
at both energies.
The dijet selection criteria resulted in $ 1.1 \times 10^4$ ($ 1.2 \times 10^4$) events in the $\sqrt{s}$ = 0.9~TeV ($\sqrt{s}$ = 7~TeV) sample.

\section{Monte Carlo Models}
\label{sec:MC}
In this section, the main features of the MC models used for comparison with the data are presented, focusing on the implementation and tuning of the UE.
Several tunes of the {\sc pythia}6 (version 6.420)~\cite{p6} event generator were used,
each one providing a different description of the non diffractive component: D6T~\cite{rdf1}, DW~\cite{rdf1}, Pro-Q20~\cite{Buckley:2009bj}, Pro-pT0~\cite{Buckley:2009bj}, Z2, as well as the central Perugia 2011 tune (P11)~\cite{Skands:2010ak} and the ATLAS minimum-bias tune 1 (AMBT1)~\cite{AMBT1}.

The tunes differ in the choice of flavour, fragmentation, and UE parameters. The latter set of parameters, which are expected to be important for these measurements, includes parameters for the parton showers, cut-off values for the multiple-parton interactions, parameters determining the geometrical
overlap between the incoming protons, and probabilities for colour reconnection. Some of the tunes have common flavour and fragmentation parameters, which have been determined using data from LEP. With the exception of the parameter settings in Z2, P11, and AMBT1, which are based on measurements from the LHC, the other {\sc pythia}6 UE tunes were
determined using data
from the Tevatron. Tune Z2 is almost identical to tune Z1~\cite{Field:Z1}, only differing in the choice of parton distribution functions. An overview of the tunes can be found in Ref.~\cite{Skands:2010ak}.

The P11 and Z2 tunes, as well as {\sc pythia}8~\cite{Pythia8},
use a new model~\cite{Skands_Wicke:2007} where multiple-parton interactions are  interleaved with parton showering.
The parton distribution functions used are the CTEQ6L~\cite{CTEQ6} set for D6T and Z2, and the CTEQ5L~\cite{CTEQ5} set for the remaining tunes. Hadronisation in {\sc pythia} is based on the Lund string fragmentation model~\cite{Lund1}. Predictions obtained from {\sc pythia}8 correspond to the default version of the generator, i.e.,without prior tuning to data.
In contrast to {\sc pythia}6, in {\sc pythia}8 the multiple-parton interactions, initial-state and final-state radiation are all interleaved, and hard diffraction has been included.

The {\sc herwig++} (version 2.5)~\cite{Bahr:2008pv} MC event generator was also used for comparisons with data. It is based on matrix-element calculations similar to those used in  {\sc pythia}. In both generators, the evolution of the parton distributions functions with momentum scale is driven by the DGLAP~\cite{\DGLAP} equations. However, {\sc herwig++} features angular-ordered parton showers and uses cluster fragmentation for the hadronisation. {\sc herwig++}  has been tuned independently for different centre-of-mass energies. Parameters for the UE and colour reconnections were tuned to UE and minimum-bias data at $\sqrt{s}$ = 0.9~TeV (tune MU900-1) and to UE data at $\sqrt{s} =$ 7~TeV (tune UE7-1)~\cite{releasenotes}.

Predictions from the event generator {\sc cascade}~\cite{Jung:2001,Jung:2010si} are compared to data from the dijet event samples. In contrast to {\sc pythia}6 and {\sc herwig++}, {\sc cascade} is based on the CCFM~\cite{\CCFM} evolution equation for the initial-state cascade, supplemented with off-shell matrix elements for the hard scattering. Multiple-parton interactions are not implemented in {\sc cascade}.

The {\sc dipsy} MC event generator~\cite{dipsy} is based on a dipole picture of BFKL~\cite{\BFKL} evolution. It includes multiple-parton interactions and can be used to predict non diffractive final states. Currently, {\sc dipsy} is not tuned to experimental data.

Data are also compared to predictions obtained from \Pp\Pp\ MC event generators used in cosmic-ray physics \cite{d'Enterria:2011kw}.
 {\sc epos}1.99~\cite{epos}, {\sc qgsjet}II~\cite{QgsjetII}, {\sc qgsjet}01~\cite{qgsjet01}, and {\sc sibyll}~\cite{sibyll} are considered. All models take into consideration contributions from both soft- and hard-parton dynamics.
In general the soft component is described in terms of the exchange of virtual quasi-particle states (a Pomeron at high energies) as in Gribov's Reggeon field theory~\cite{Gribov:1968fc}. At higher energies and scales, the interaction is described by perturbative QCD (with DGLAP evolution)  by an extension of the approach used to describe the soft region.
The models were not tuned to LHC data.

\section{Corrections}
\label{sec:Corrections}
The data were corrected to the stable-particle ($\tau >  10^{-12}$~s)  level
with the help of bin-by-bin correction factors derived from simulated events generated with the {\sc pythia}6.4 MC event generator  and passed through the CMS detector simulation based on {\sc Geant4}~\cite{Agostinelli:2002hh}.
The position and width of the beam spot in the simulation were adjusted to that determined from the data. The simulated events were processed using the same analysis chain as for collision data. Corrections to the measured data
were applied bin-by-bin to account for acceptance, inefficiency, and bin-to-bin migrations due to the detector resolution.

The correction factor for each bin was determined
as the average of the correction factors from the {\sc pythia}6 P0~\cite{Skands:2010ak} D6T, DW, Z2, and Pro-Q20 tune predictions. All the tunes described reasonably well the shape of the data before the correction. The deviation from the average was included in the model-dependent systematic uncertainty. The number of selected events in each of the various MC samples
were comparable with the data set size.

The correction factors were calculated as the ratio of MC predictions at the stable-particle level and the detector level.
At the stable-particle level, the selected particles were required to be in the same pseudorapidity range as that used for the measurements (3.15$<|\eta|<$4.9), without any further requirement on their energies. Neutrinos and muons were excluded. In addition, at least one charged particle was required, on each side, within the acceptance of the BSC ($3.9 <|\eta|< 4.4$). This requirement was imposed in order to replicate
the use of the BSC triggers in the detector-level analysis chain.

For the analysis of dijet events, the kinematic selection of the dijet system was the same for the stable-particle level and detector-level jets, i.e.,~$|\eta_{jet}| < 2.5$, $|\Delta\phi(jet_{\textrm{1}},jet_{\textrm{2}})-\pi| < 1.0$, and $p_{T,jet} > 8$~GeV ($p_{T,jet}> 20$~GeV) for the $\sqrt{s} = 0.9$~TeV ($\sqrt{s}  = 7$~TeV) analysis.

The event selection criteria applied at the stable-particle level in the MC simulation are summarized in Table~\ref{tab_selection}.

The bin-by-bin correction factors range from 1.6 to 2.0 (1.5 to 2.0) for the $\sqrt{s} = 0.9$~TeV ($\sqrt{s} = 7$~TeV) minimum-bias and dijet data in the  four lowet  $|\eta|$ bins of the measurement. Because of the larger amount of dead material within the range of the highest-$|\eta|$ bin, the correction factor is larger, 2.5 and 2.2, for the  $\sqrt{s} = 0.9$~TeV and $\sqrt{s} = 7$~TeV measurements, respectively. The nonlinear behaviour of the HF calorimeters was implemented in the detector simulation and was therefore taken into account in the bin-by-bin correction.

\begin{table*}[hbH]
\begin{center}
\caption{Event selection criteria applied at the stable-particle level in the MC simulation.
 \label{tab_selection}}
\begin{tabular}{|c|}
\hline
Minimum-bias event selection  \\
\hline
 $N_{\textrm{charged particles}} >0 $ in  $3.23 <\eta< 4.65 \textrm{ and}  -4.65<\eta<-3.23  $   \\
\hline
\hline
Dijet event selection  \\
\hline
  $N_{\textrm{charged particles}} >0 $ in  $3.23 <\eta< 4.65 \textrm{ and}  -4.65<\eta<-3.23  $   \\
$|\eta_{jet}| <  2.5$\\
$|\Delta\phi(jet_{\textrm{1}},jet_{\textrm{2}})-\pi| < 1.0$\\
$p_{T,jet} > 8$~GeV ($\sqrt{s} = 0.9$~TeV) \\
$p_{T,jet}> 20$~GeV ($\sqrt{s} = 7$~TeV)  \\
\hline
\end{tabular}
\end{center}
\end{table*}

\section{Systematic Uncertainties}
\label{sec:Systematics}
\begin{table*}[hbH]
\begin{center}
\caption{Systematic uncertainties on the energy-flow measurement for the minimum-bias and dijet analyses and their sum. The ranges indicate the variation of the uncertainty within the different $|\eta|$ bins. \label{tab_syst}}
\begin{tabular}{|c|c|c|}
\hline
Effect & Minimum-bias analysis & Dijet analysis  \\
\hline
Energy scale & 10\% & 10\%  \\
Channel-to-channel miscalibration & 1\% & 1\%  \\
Minimum energy in calorimeter towers & 2\% & 2\%  \\
Photomultiplier hits rejection algorithm & 3\% & 3\%  \\
Primary vertex $z$ position & 1\% & 1\%  \\
Model uncertainty, $\sqrt{s}$= 0.9 / 7 TeV & 1--3\% /  1--2\%  & 4--11\% / 12--17\%  \\
Non uniformity corrections & 0--3\% & 0--3\%  \\
Short-fibre response, $\sqrt{s}= 0.9$ / 7 TeV & 3--9\% /  3--6\%  & 6--18\% / 6--8\%  \\
Jet energy scale & - & 2\%  \\
\hline
Total, $\sqrt{s} = 0.9$~TeV (7~TeV)  & 11--15\% (12--13\%) & 13--24\% (17--22\%) \\
\hline
\end{tabular}
\end{center}
\end{table*}

The systematic uncertainties on the energy-flow measurement  are summarised in Table~\ref{tab_syst}. The total uncertainty for each $|\eta|$ bin was obtained by adding all the uncertainties in
quadrature. Depending on the centre-of-mass energy and the event selection, the total systematic uncertainty was found to be 11 -- 15\% and 13 -- 22\% for the minimum-bias and dijet analyses, respectively. An important systematic effect in the measurement of the forward energy flow is the global energy-scale uncertainty of the HF calorimeters,
which was estimated to be 10\% using $Z\to e^+e^-$ events with one electron in the HF calorimeter and the other reconstructed in other subdetectors.

To estimate the effect of channel-to-channel miscalibration of the HF calorimeters, the response per channel was randomly varied between $\pm 10\%$. The resulting energy flow was shifted by less than 1\%. To estimate the possible influence of any remaining calorimeter noise, the requirement on the minimum energy deposit in a tower was increased to 4.5~GeV. This resulted in a change in the energy-flow distributions by less than 2\%.

Variations of the algorithms used to reject events with abnormal signals in the HF calorimeters, due to
particles hitting the photomultipliers,
give a systematic uncertainty on the energy flow of approximately 3\%.

The pseudorapidity $\eta$ is defined with respect to the origin of the CMS reference coordinate system. For events with a primary vertex far (e.g.,~$|z| > 15$\unit\cm) from that point, the distributions of measured variables as a function of $\eta$ are shifted. To estimate the influence of this effect, the energy flow was calculated separately for events with the primary vertex restricted to the ranges $|z|<$ 4 \unit\cm, $4 <  |z|  < 9$  \unit\cm and $9 < |z| < 15$ \unit\cm. The largest difference was approximately 1\%.

The systematic uncertainties due to the model dependence of the bin-by-bin corrections were estimated from the differences between  the correction factors obtained from the various MC tunes and the average value. The differences were applied to the data points uniformly. Depending on the centre-of-mass energy and the $|\eta|$ bin, the resulting changes were between 1 and 17\%.
Possible residual systematic effects resulting from the non linearity of the HF calorimeters arise from the difference in the particle spectra and energy distributions predicted by the various MC tunes. This was also included in the model dependence of the correction factor.

The uncertainty on the non uniformity corrections leads to a systematic uncertainty in the energy flow of approximately 3\% for the highest $|\eta|$ bin, and is negligible for the lowest bin.

An additional systematic uncertainty comes from the simulation of the short-fibre response in the HF, determined by
repeating the full analyses using the long fibres exclusively.
The former
was not fully described in the HF simulation at low energies. The resulting uncertainty on the energy-flow measurement was between 3 and 9\% for the minimum-bias datasets. For the dijet samples the uncertainty was between 6 and 10\%, except for the two highest-pseudorapidity bins for $\sqrt{s} = 0.9$~TeV, where it was between 13 and 18\%.

In the case of the dijet data sample, an additional uncertainty arises from the jet energy scale. This was estimated by varying the jet energy by $\pm$10\%, leading to an uncertainty in the energy flow of 2\%.

The contribution of beam-gas and non interaction events to the event sample was investigated by performing the minimum-bias selection on events triggered when there was no beam crossing. It was found that no such event passed the selection criteria. Other sources of systematic effects such as pile up and diffractive modelling were each found to contribute less than 1\% to the total systematic uncertainty.

\section{Results}
\label{sec:Results}
\begin{figure}[htbp]
\begin{center}
\includegraphics[width=0.49\textwidth]{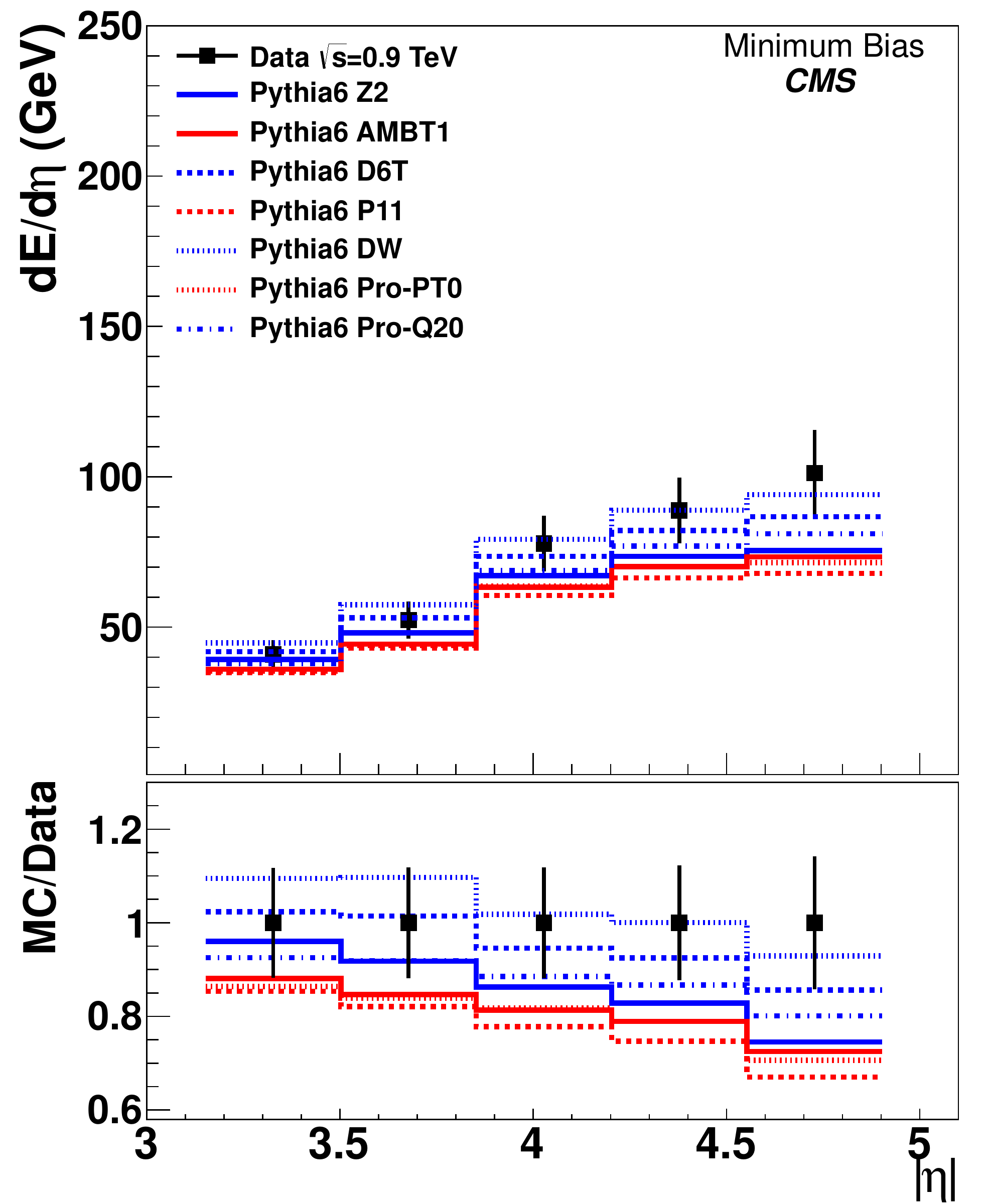} \hspace{0.0cm} \includegraphics[width=0.49\textwidth]{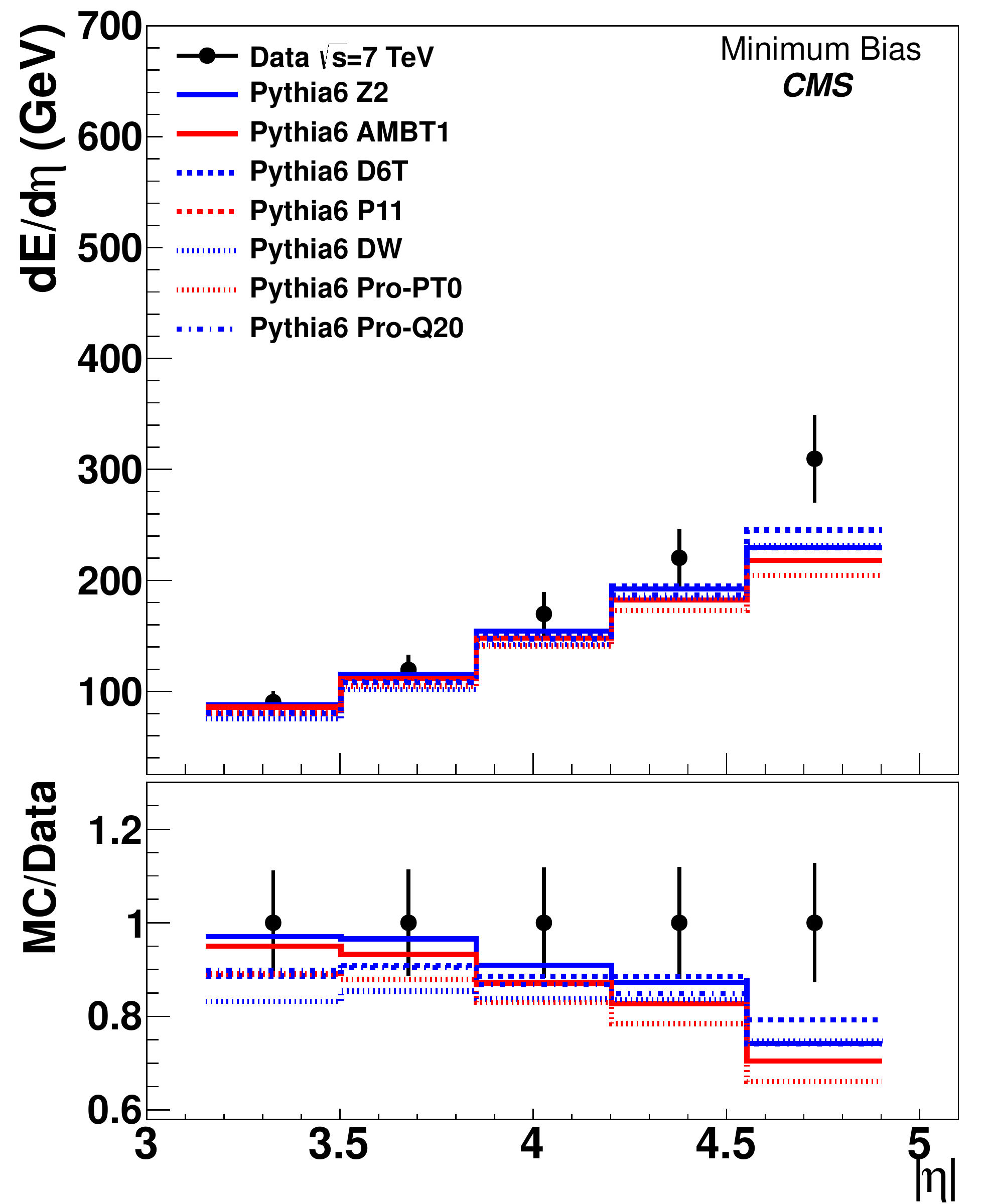}
\includegraphics[width=0.49\textwidth]{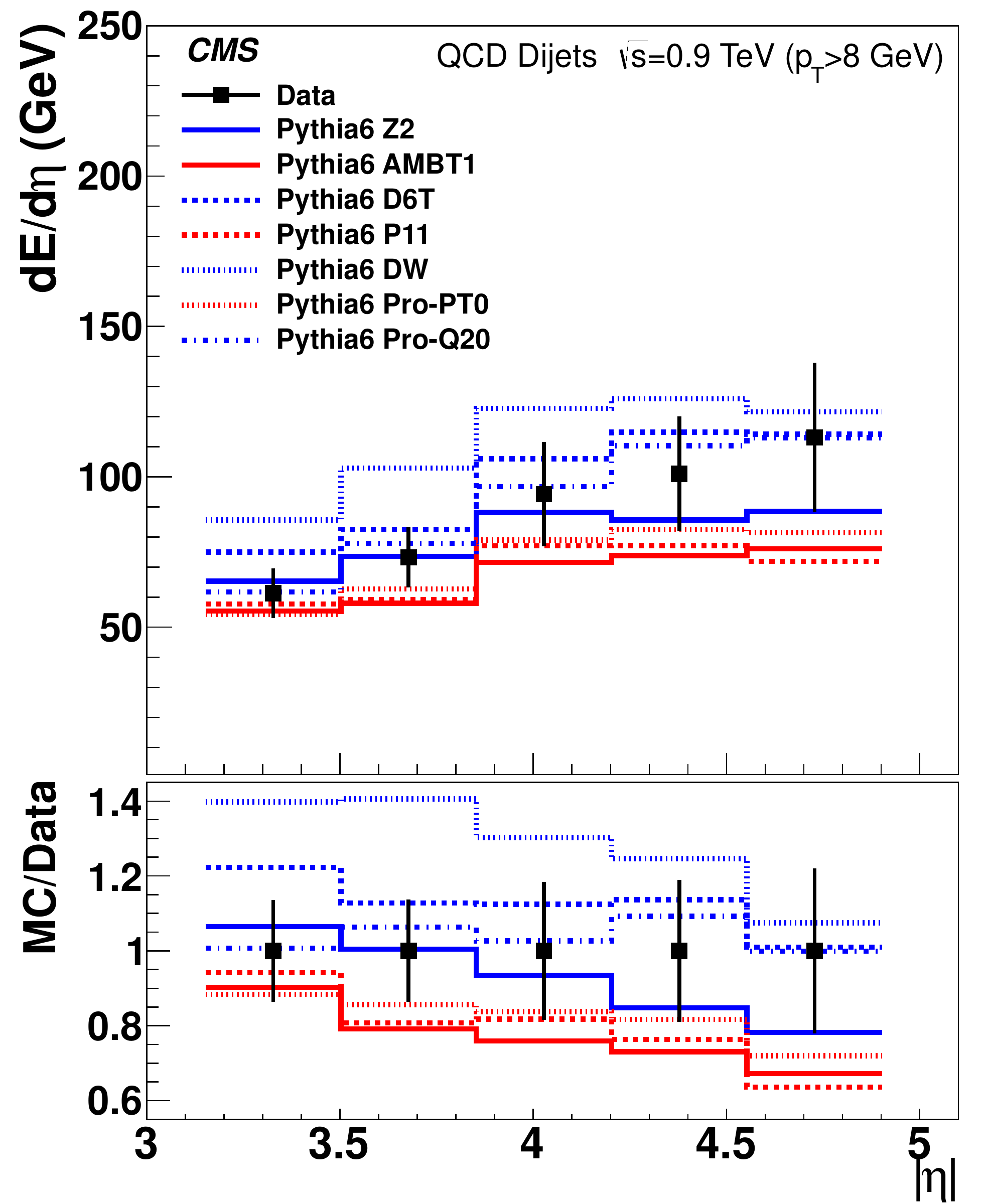} \hspace{0.0cm} \includegraphics[width=0.49\textwidth]{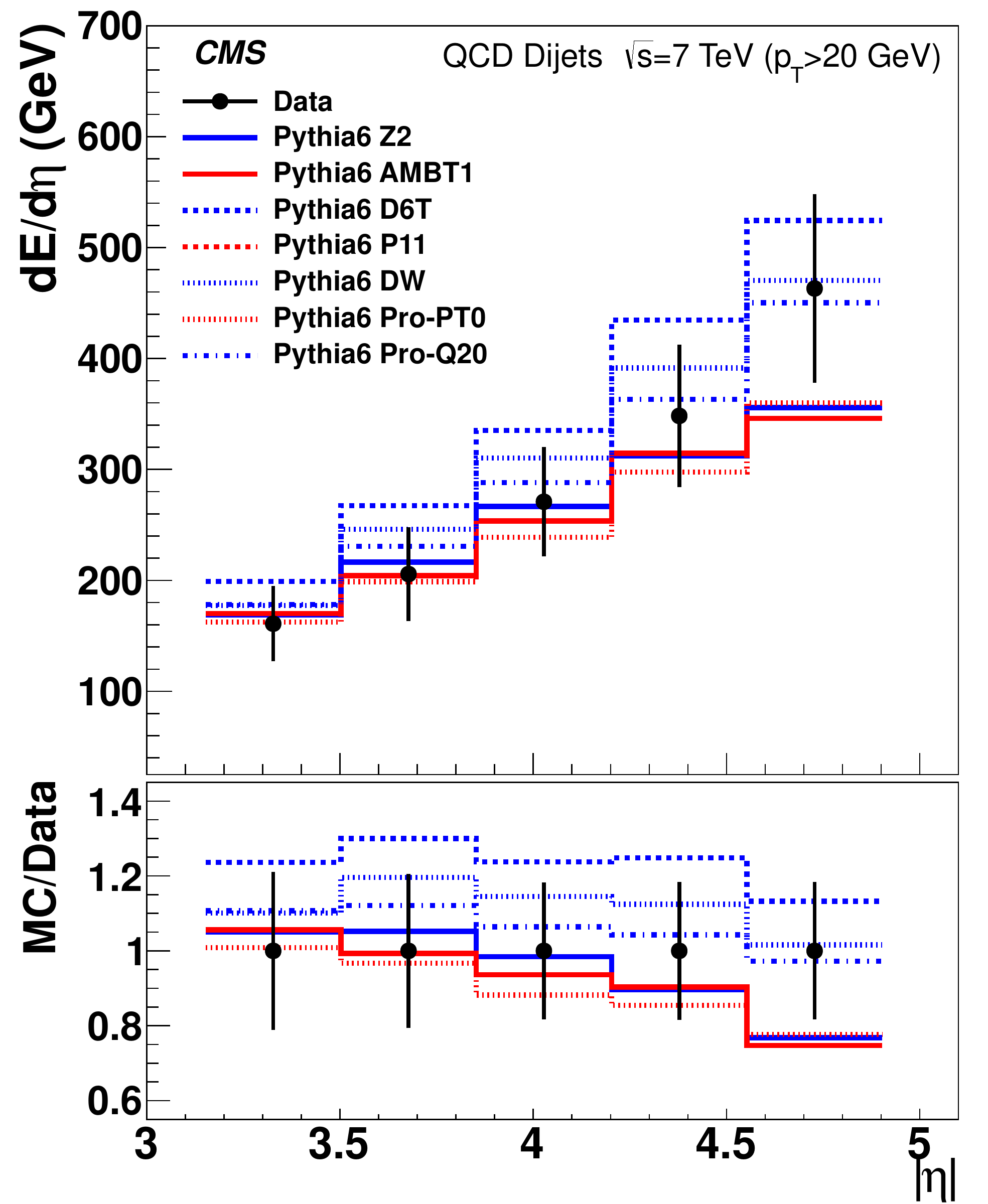}
\caption{Energy flow  as a function of $\eta$ for minimum-bias (upper) and dijet (lower) events at $\sqrt{s}  = 0.9$ and 7~TeV. The data are shown as points with error bars, while the histograms correspond to predictions obtained from various {\sc pythia}6 tunes. The error bars represent the systematic uncertainties, which are strongly correlated between the bins. The statistical uncertainties are negligible.
The lower panels show the ratio of MC prediction to data.
}
\label{fig:results_mb_mctunes}
\end{center}
\end{figure}

The energy flow was measured with the CMS HF calorimeters at large pseudorapidities, $3.15  <|\eta|< 4.9$, and corrected to the stable-particle level. The results are shown in Fig.~\ref{fig:results_mb_mctunes} for minimum-bias and dijet events, at  $\sqrt{s} = 0.9$~TeV and $\sqrt{s} = 7$~TeV. The systematic uncertainties are indicated as error bars; they are correlated between the $|\eta|$ bins. The statistical errors are negligible. The data are also presented in Tables~\ref{tab:mbdata} and~\ref{tab:dijetdata}.

We observe three distinct features of the data. The first is that the energy flow in both minimum-bias
and dijet events increases with pseudorapidity, and the increase is found to be steeper for minimum-bias events. The second is
that the energy flow increases with centre-of-mass energy, being
a factor of two to three
higher at  $\sqrt{s} = 7$~TeV than at $\sqrt{s} = 0.9$~TeV.
The increase in energy flow for minimum-bias events is larger than what is
observed in the charged-particle multiplicity, $dN_{ch}/d\eta$, reported in Ref.~\cite{cms_dndeta}. Finally, the average energy flow is significantly higher
in dijet events than in the minimum-bias sample.

The data are compared to different MC predictions.
In Fig.~\ref{fig:results_mb_mctunes} (upper) the minimum-bias measurements are compared to the predictions of various {\sc pythia}6 tunes. An interesting observation is the similarity between the predictions of tunes Z2 and AMBT1, both based on LHC data, and the older D6T tune. Predictions obtained from several other {\sc pythia}  tunes  are also shown. The variation of the predictions is on the order of 10--20\%.

\begin{figure}[htbp]
\begin{center}
\includegraphics[width=0.49\textwidth]{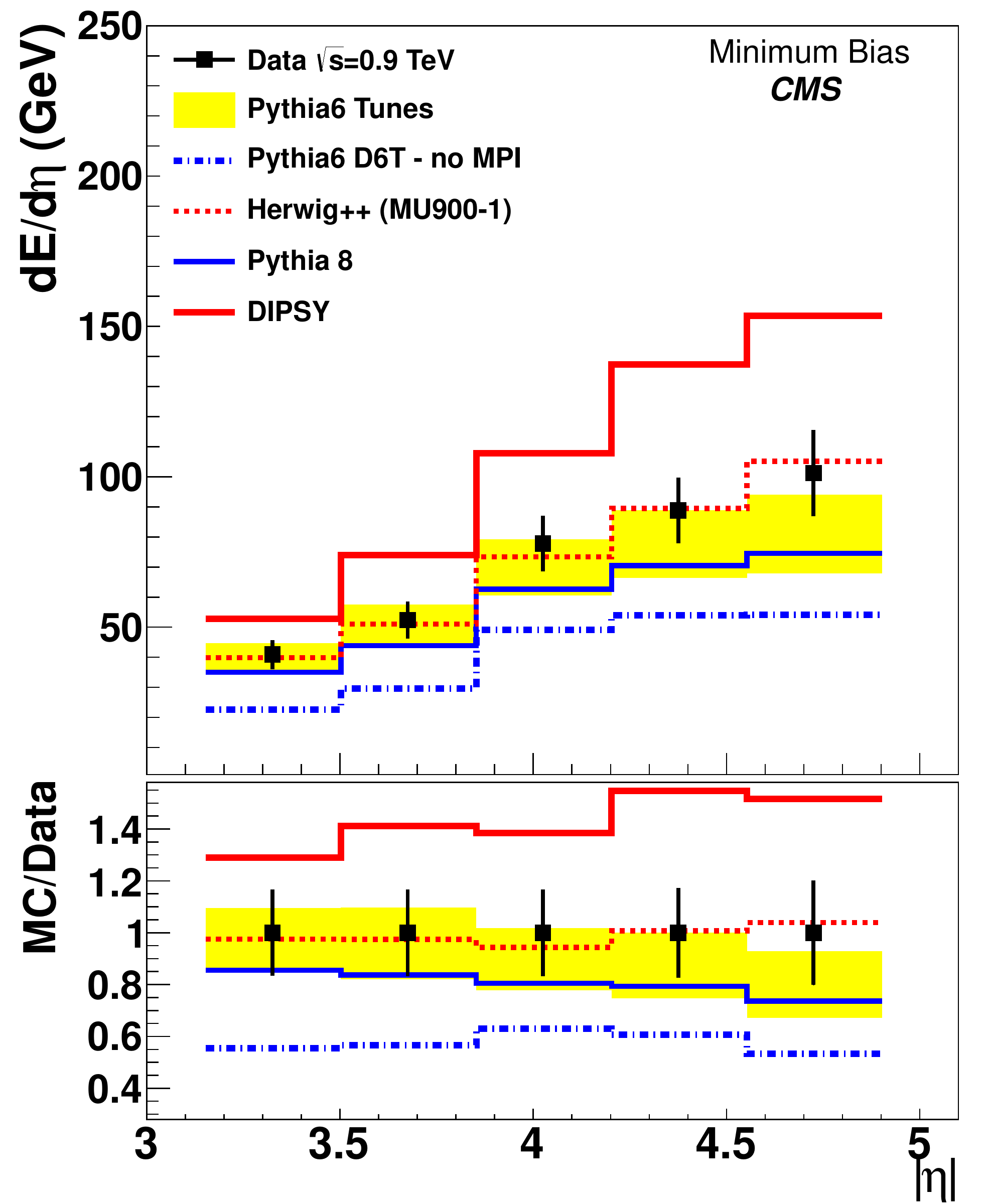} \hspace{0.0cm} \includegraphics[width=0.49\textwidth]{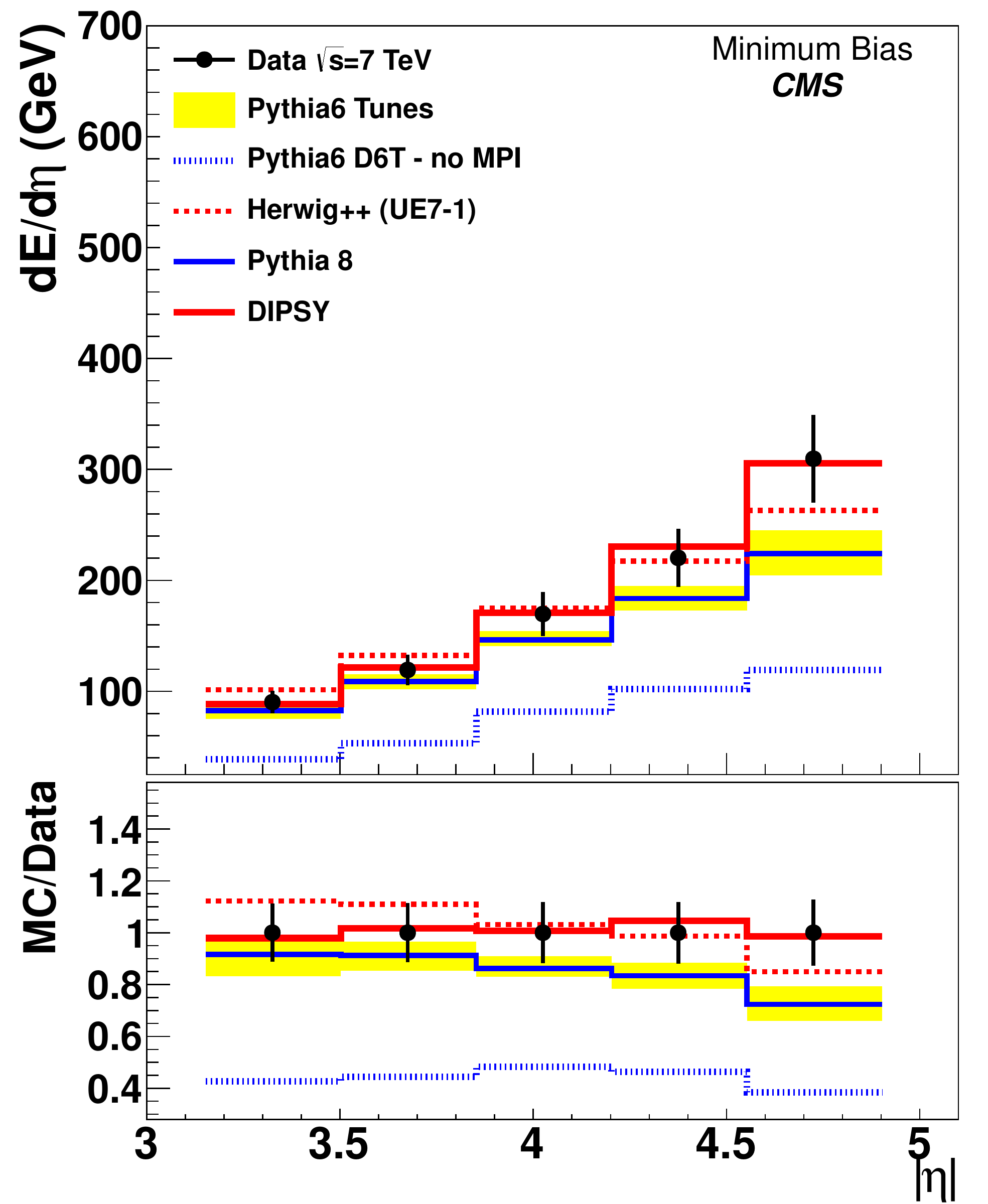}
\includegraphics[width=0.49\textwidth]{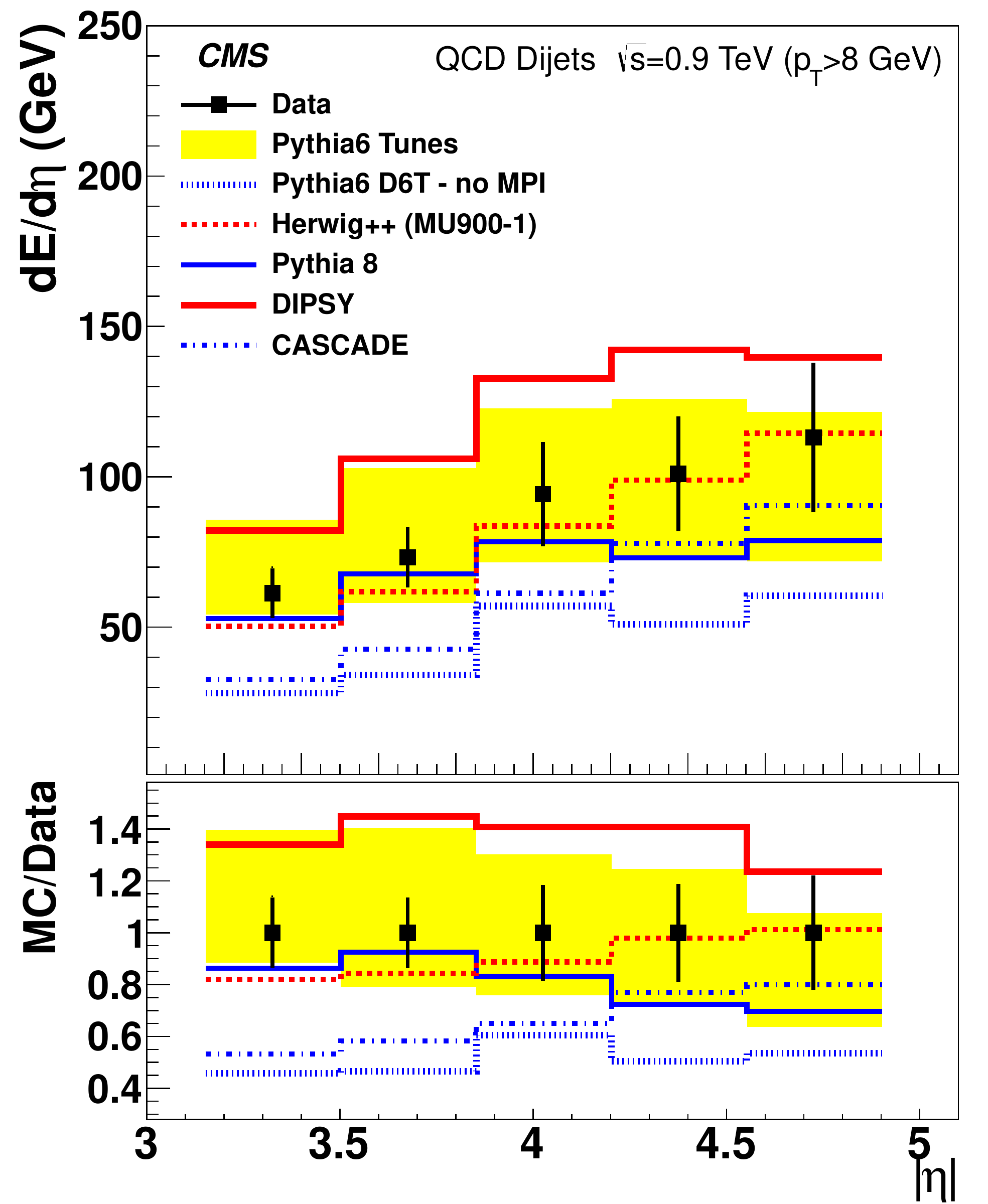} \hspace{0.0cm} \includegraphics[width=0.49\textwidth]{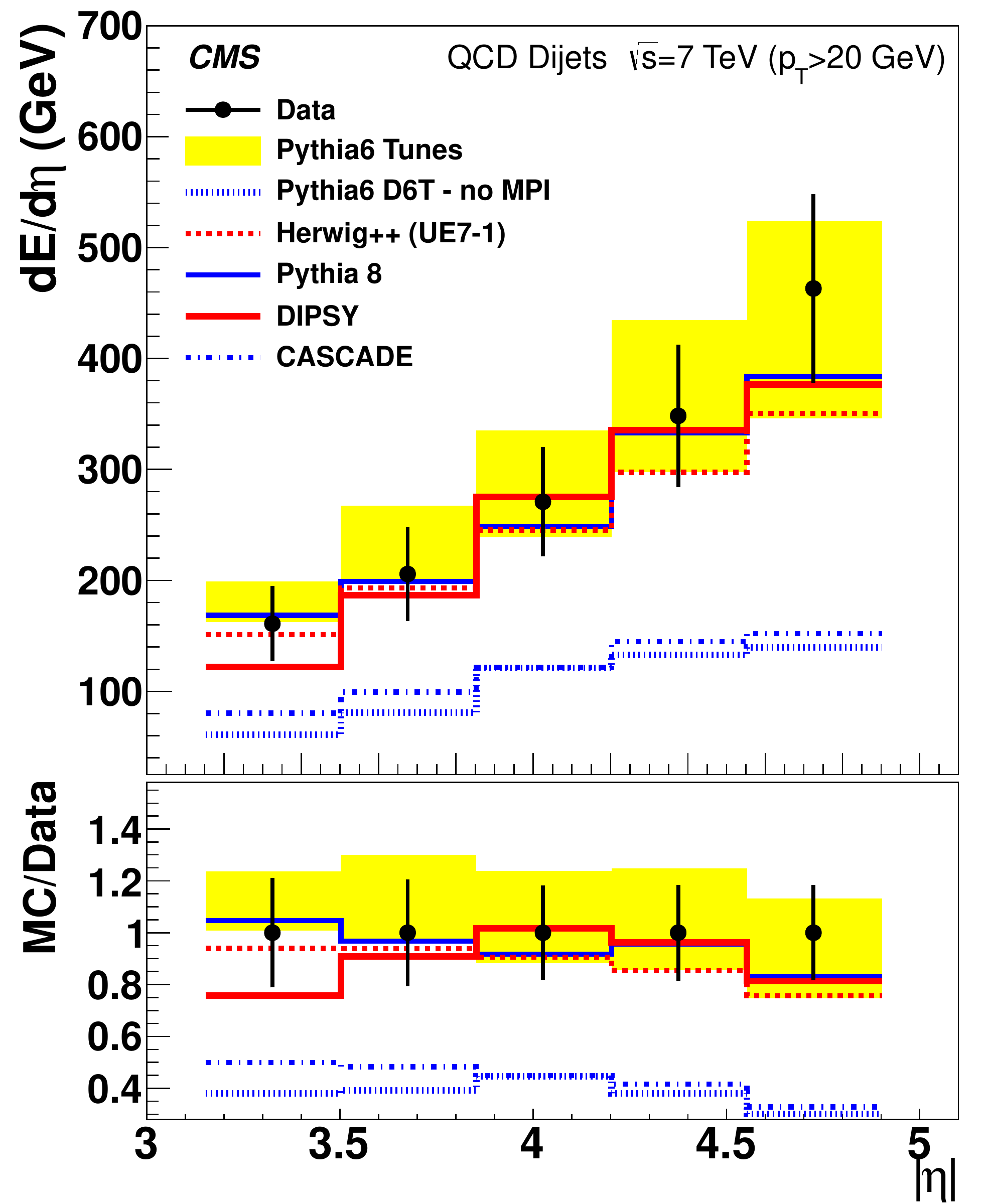}
\caption{Energy flow  as a function of $\eta$ for  minimum-bias (upper) and dijet (lower) events at $\sqrt{s} = 0.9$~TeV and $\sqrt{s} = 7$~TeV. The data are shown as points with error bars, while the histograms correspond to predictions obtained from various Monte Carlo event generators. The yellow bands illustrate the spread of the predictions from the different {\sc pythia}6 tunes considered. The bands are obtained by taking the minimum and maximum variations of the {\sc pythia}6 tunes shown in fig.~\ref{fig:results_mb_mctunes}. The predictions from {\sc herwig++} are made with tunes specific to the respective centre-of-mass energy. The error bars represent the systematic uncertainties, which are strongly correlated between the bins. The statistical uncertainties are negligible.
The lower panels show the ratio of MC prediction to data.
}
\label{fig:results_mb_gens}
\end{center}
\end{figure}
In Fig.~\ref{fig:results_mb_gens} (upper) the minimum-bias measurements are compared to results from different Monte Carlo event generators.
The {\sc pythia}6 tunes (also shown in Fig.~\ref{fig:results_mb_mctunes}) are presented as a band, which is constructed from the minimum and maximum values of the different {\sc pythia}6 predictions in each bin. The tunes giving the limits of the bands are different for each bin.
Also shown are the predictions of {\sc pythia}8, {\sc herwig++},{\sc dipsy}, and {\sc pythia}6 D6T without multiple-parton interactions.
The prediction from {\sc pythia} without multiple-parton interactions is at least 40\% lower than the measurement. We checked that the same prediction at the parton level (without hadronisation) gives smaller energy flow, by approximately 20\%. We also found that changing the maximum value for the scale used in the parton shower
 by a factor of four increases the energy flow only by $\sim$5\%. Therefore, these effects cannot bring the prediction without multiple-parton interactions into agreement with the data.
The {\sc herwig++} tunes describe the measurements well at both centre-of-mass energies. The {\sc pythia}8 predictions are always within the tune uncertainty band of {\sc pythia}6 and give a slightly flatter energy flow distribution than the data. {\sc dipsy}, without prior tuning, describes the minimum-bias data  well for $\sqrt{s} = 7$~TeV. However, it overestimates the energy flow for $\sqrt{s} = 0.9$~TeV, by up to 50\%.
\begin{figure}
\begin{center}
\includegraphics[width=0.49\textwidth]{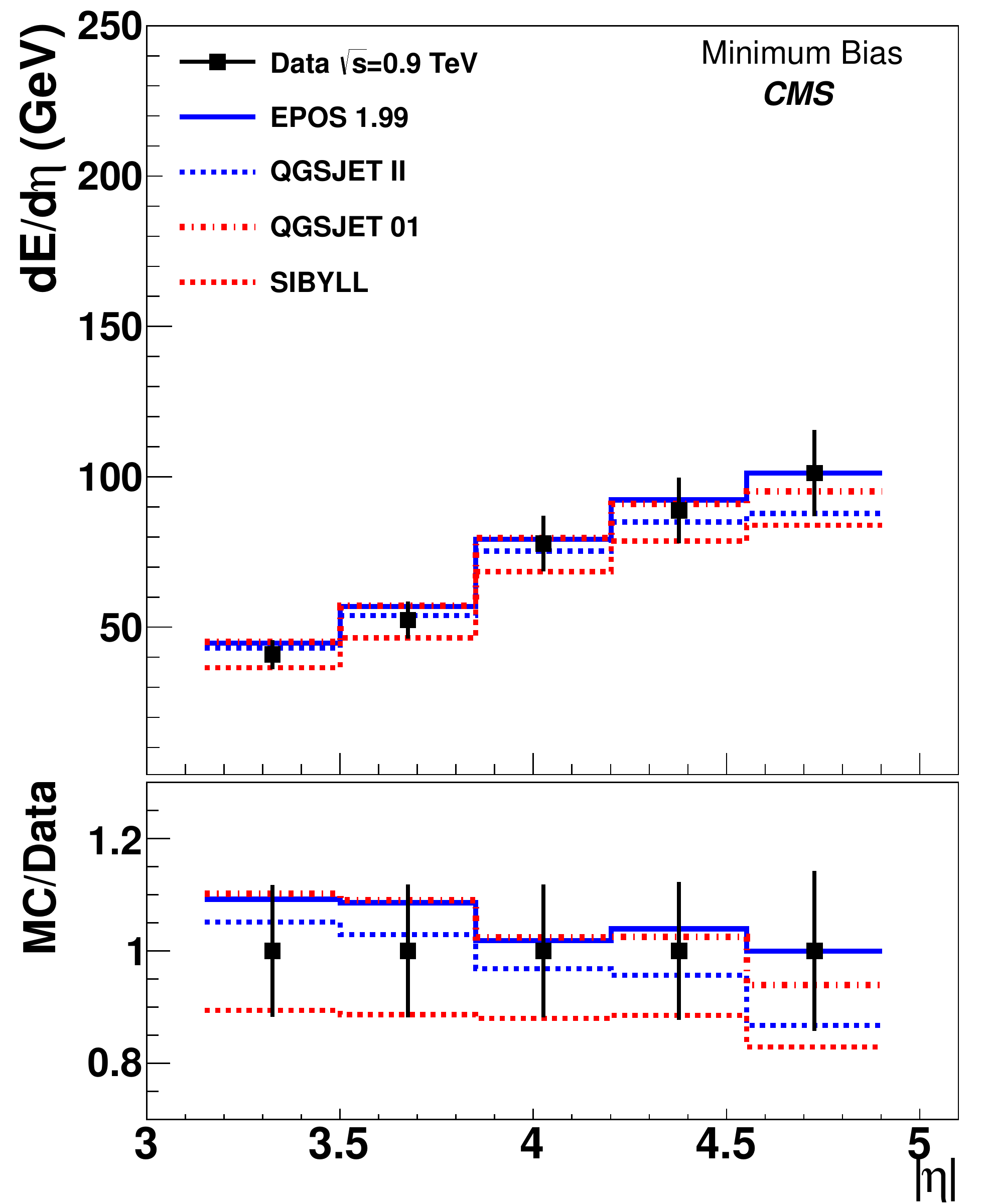} \hspace{0.0cm} \includegraphics[width=0.49\textwidth]{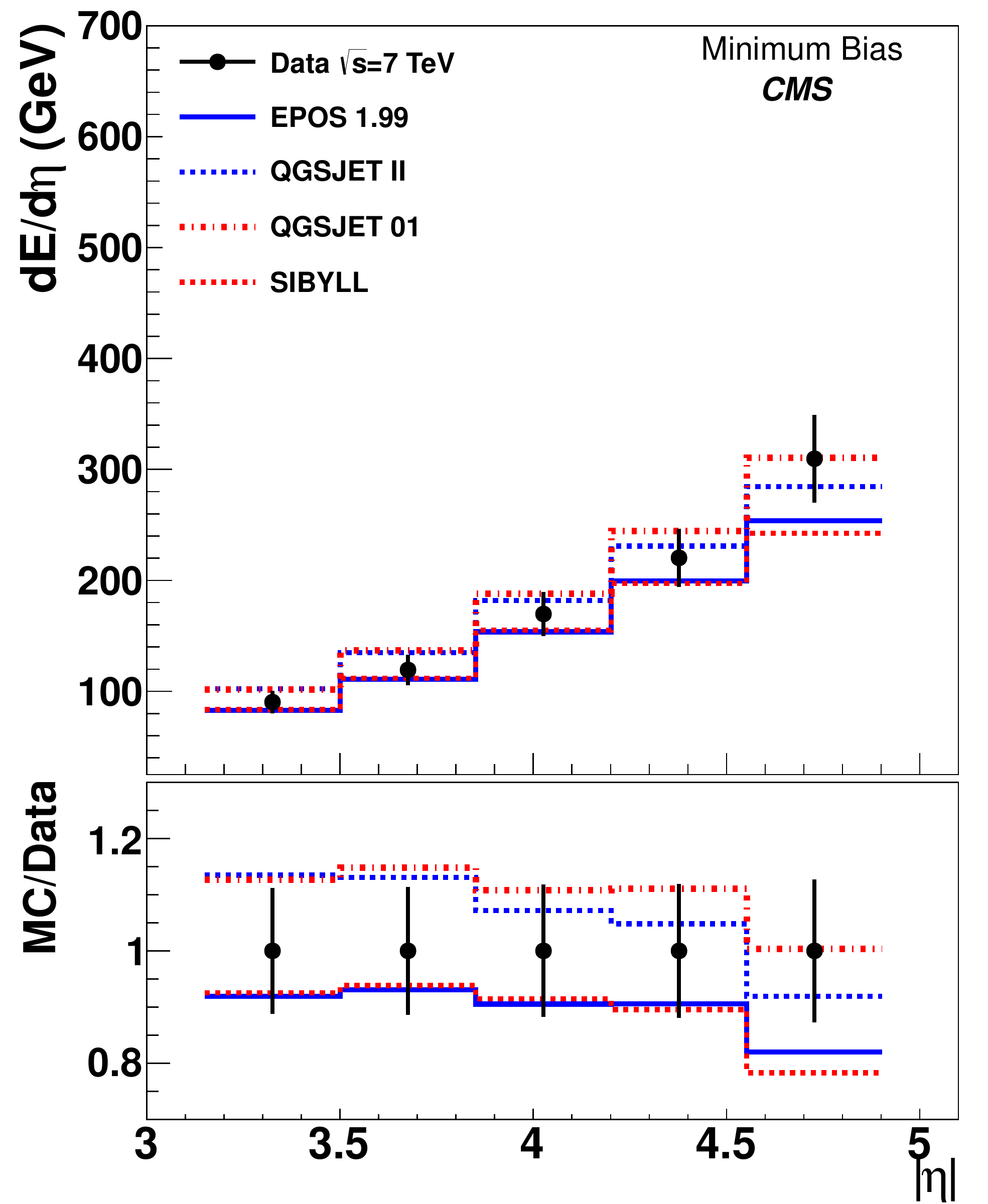}
\includegraphics[width=0.49\textwidth]{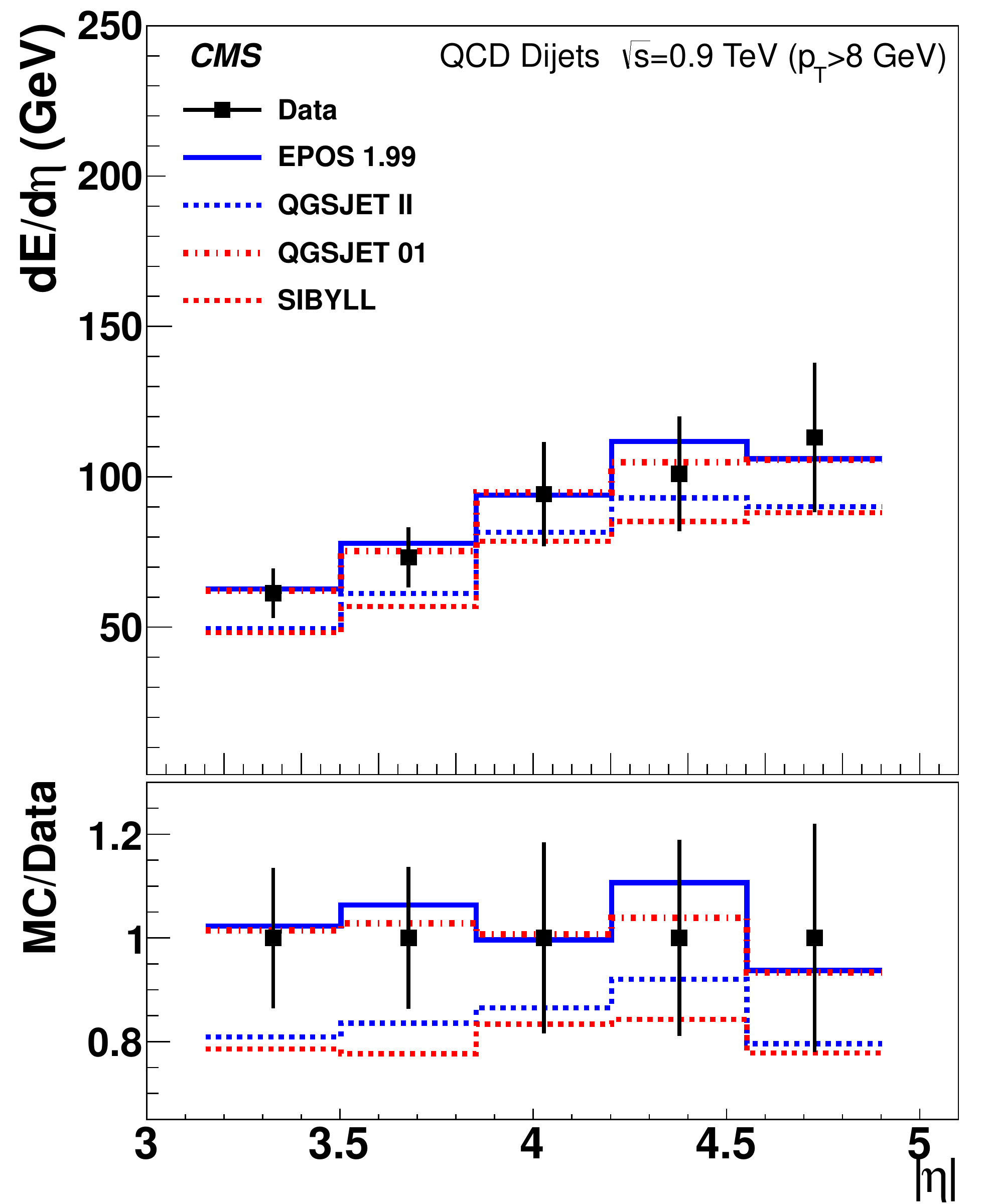} \hspace{0.0cm} \includegraphics[width=0.49\textwidth]{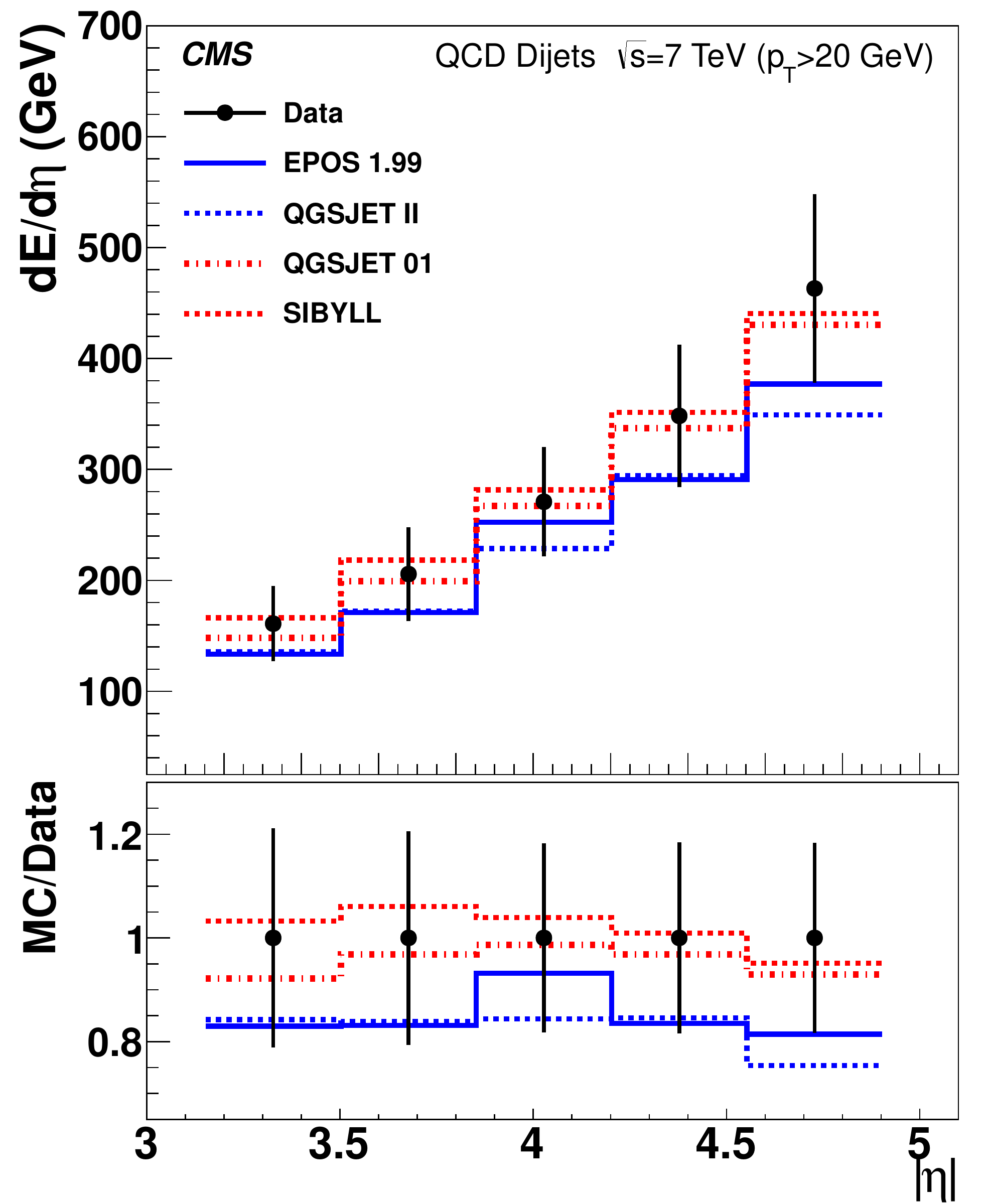}
\caption{Energy flow as a function of $\eta$ for minimum-bias (upper) and dijet (lower) events at $\sqrt{s} = 0.9$~TeV and $\sqrt{s} = 7$~TeV. The data are shown as points with error bars, while the histograms correspond to predictions obtained from different cosmic-ray Monte Carlo event generators. The error bars represent the systematic uncertainties, which are strongly correlated between the bins. The statistical uncertainties are negligible.
The lower panels show the ratio of MC prediction to data.
}
\label{fig:results_mb_cosmic}
\end{center}
\end{figure}
In Fig.~\ref{fig:results_mb_cosmic}  the measured energy flow is compared to predictions derived from the event generators {\sc epos}, {\sc qgsjet}II,  {\sc qgsjet}01, and {\sc sibyll}, which are used in cosmic-ray air shower simulations. The description of the minimum bias data by all these models is good.

The results for the energy flow in dijet events are shown in Figs.~\ref{fig:results_mb_mctunes}--\ref{fig:results_mb_cosmic} (lower). The data are the same in the three figures. In general, the rise of the average energy with increasing pseudorapidity is described by all the models, and the agreement with the data is better than in the case of minimum-bias events.
The predictions from {\sc pythia} with different parameter settings for multiple-parton interactions agree with the data, whereas the prediction without multiple-parton interactions is too low. The prediction from {\sc cascade}, which uses a different parton shower evolution, is also too low compared to the data, although slightly higher than the prediction from {\sc pythia} without multiple-parton interactions.
The energy flow obtained from {\sc herwig++} is in good agreement with the measurement, while that from {\sc dipsy} is consistent with the data at $\sqrt{s} = 7$~TeV, but too high at $\sqrt{s} = 0.9$~TeV. The predictions from the  cosmic-ray MC generators are very close to the data, as shown in Fig.~\ref{fig:results_mb_cosmic} (lower). However, a larger deviation is observed in the dijet measurement at $\sqrt{s} = 0.9$~TeV, where {\sc qgsjet}II and {\sc sibyll} underestimate the data in the lowest $|\eta|$ bins, while {\sc qgsjet}01 is in agreement with the measurement.

We have checked that the energy flow in dijet events, after subtracting the energy flow from minimum-bias events, is still significantly larger than that predicted from MC models without multiple-parton interactions. The disagreement is found to be at least a factor of three. This suggests that the energy flow in dijet events is composed of more than a soft underlying event and a single parton interaction with a parton shower.

\begin{figure}
\begin{center}
\includegraphics[width=0.49\textwidth]{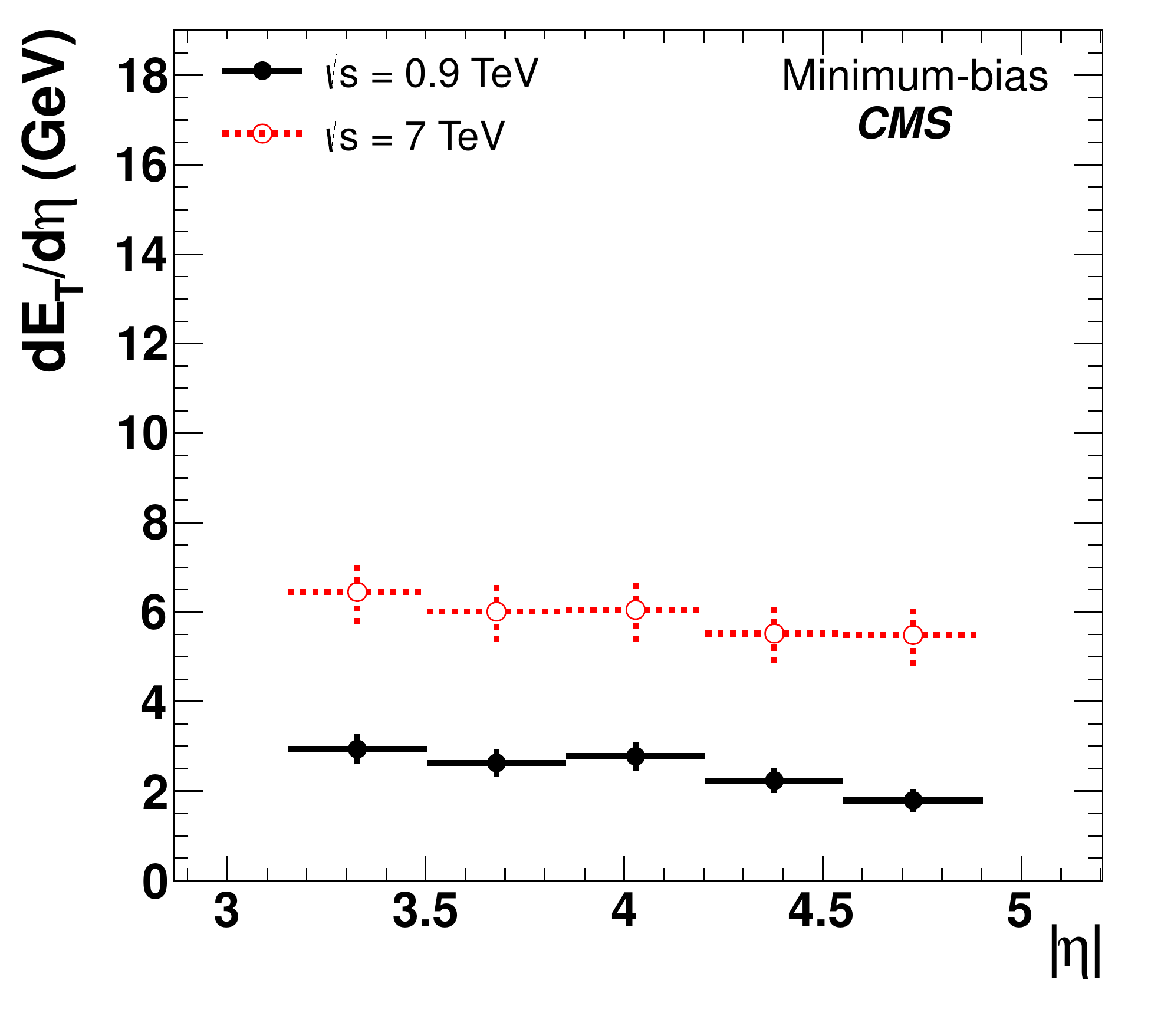} \hspace{0.0cm} \includegraphics[width=0.49\textwidth]{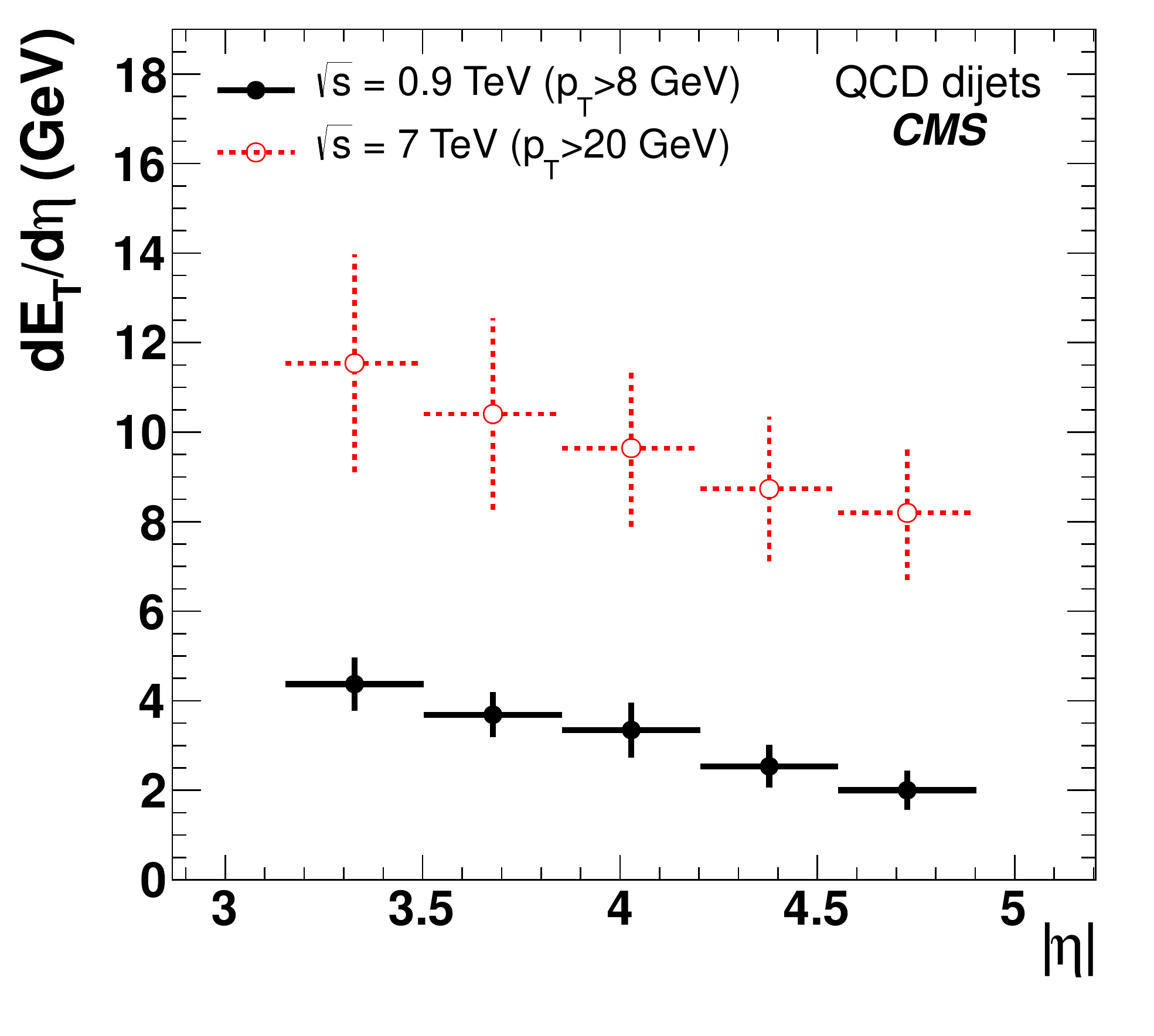}
\caption{Transverse energy flow as a function of $\eta$ for minimum-bias and dijet  events at $\sqrt{s} = 0.9$~TeV and $\sqrt{s} = 7$~TeV. The data are shown as points with error bars. The error bars represent the systematic uncertainties, which are strongly correlated between the bins. The statistical uncertainties are negligible.
}
\label{fig:results_Et}
\end{center}
\end{figure}

From the measured energy flow, we can estimate the transverse energy per pseudorapidity bin~$i$, given by
$ E_{T\,i} = E_i \sin\theta_i$.
The increase of the energy flow with increasing pseudorapidity in the minimum-bias events leads to a constant average transverse energy
of $d E_T/d \eta \approx 3 $~GeV at $\sqrt{s} = 0.9$~TeV rising to $d E_T/d \eta \approx 6 $~GeV at $\sqrt{s} = 7$~TeV.
The measured transverse energy as a function of pseudorapidity is shown in  Fig.~\ref{fig:results_Et}. The error bars were obtained by propagating the systematic uncertainties of the energy-flow measurements.

In dijet events the average energy also increases with increasing pseudorapidity, but the increase is less steep than in minimum-bias events, leading to a distribution in $E_T$ that decreases at  $\sqrt{s} = 0.9$~TeV from $ d E_T/d \eta \approx 4 $~GeV at $|\eta| = 3.3$ to  $  \approx 2 $~GeV at $|\eta| = 4.7$. At $\sqrt{s} = 7$~TeV the transverse energy decreases from $ d E_T/d \eta \approx 11.5 $~GeV at $|\eta| = 3.3$ to  $ \approx 8 $~GeV at $|\eta| = 4.7$ (Fig.~\ref{fig:results_Et}). This decrease of transverse energy is
consistent with
calculations using a $p_T$-ordered or virtuality-ordered ($Q^2$-ordered) parton evolution, with the highest transverse parton momentum being closest to the hard scatter (at small $|\eta|$), and decreasing towards the direction of the proton (large $|\eta|$). The general behaviour of the energy, as well as the transverse energy flow, is well described by MC models applying a $p_T$- or virtuality-ordered parton shower with multiple-parton interactions.

The average transverse energy measured in dijet events can also be compared to the transverse energy in deep-inelastic scattering (DIS) events measured at HERA~\cite{Adloff:1999ws} at a similar $x$ and $Q^2$. We compare our measurement of the transverse energy of $ d E_T/d \eta \approx 4 $~GeV at $|\eta| = 3.3$ and $\sqrt{s} = 0.9$~TeV (at a scale of $Q^2 = 4 p_T^2 \approx 250 $ GeV$^2$ and $x \sim 0.02$) with the corresponding value from DIS of $ d E_T/d \eta \approx 2  $~GeV at $x\sim 0.01$ and $Q^2 \sim 250$~GeV$^2$.
For the measurement at 7~TeV, at a scale of $Q^2 = 4 p_T^2 = 1600 $ GeV$^2$ and $x\sim 0.006$, no corresponding measurements from DIS exist. This comparison shows that the (transverse) energy flow in \Pp\Pp~collisions is significantly larger than that measured in DIS, where the contribution from multiple-parton interactions are negligible.

In summary, the shape of the energy-flow distribution, both in minimum-bias and dijet events, is reproduced by all the MC event generators that include a contribution from multiple-parton interactions. However, the magnitude of the average energy density depends significantly on the parameter settings of the different MC tunes. The measured energy flow can thus be used for further constraints on the modeling of multiple-parton interactions.

\begin{table}[htbH]
\begin{center}
\caption{Corrected energy flow $dE/d\eta$ and systematic uncertainties $\delta_{sys}$ for the minimum-bias measurements. All values are in GeV. The statistical uncertainties are less than 0.1\% in all bins, and are therefore not listed.\label{tab:mbdata}}
\begin{tabular}{|c|cc|cc|}
\hline
&\multicolumn{4}{c|}{Minimum-bias data}\\
&\multicolumn{2}{c}{$\sqrt{s} = 0.9$~TeV} & \multicolumn{2}{c|}{$\sqrt{s} = 7$~TeV}  \\
\hline
$|\eta|$  & $ dE/d\eta $  & $ \delta_{sys} $ &  $ dE/d\eta $ &  $ \delta_{sys} $  \\
\hline
3.2 - 3.5 & 41 & 5 &  90 & 10  \\
3.5 - 3.9 & 52 & 6 &  119 & 14 \\
3.9 - 4.2 & 78 & 9 &  170 & 20 \\
4.2 - 4.6 & 89 & 11 &  220 & 26 \\
4.6 - 4.9 & 101 & 14 & 310 & 40  \\
\hline
\end{tabular}
\end{center}
\end{table}

\begin{table}[htbH]
\begin{center}
\caption{Corrected energy flow $dE/d\eta$ and systematic uncertainties $\delta_{sys}$ for the dijet  measurements. All values are in GeV.  The statistical errors are less than 0.1\% in all bins, and are therefore not listed. \label{tab:dijetdata}}
\begin{tabular}{|c|cc|cc|}
\hline

&\multicolumn{4}{c|}{Dijet data}\\
&\multicolumn{2}{c}{$\sqrt{s} = 0.9$~TeV} & \multicolumn{2}{c|}{$\sqrt{s} = 7$~TeV}  \\
\hline
$|\eta|$ & $ dE/d\eta $ & $ \delta_{sys} $ &  $ dE/d\eta $ & $ \delta_{sys} $  \\
\hline
3.2 - 3.5 & 61 & 8 & 161 & 34  \\
3.5 - 3.9 & 73 & 10 & 206 & 42  \\
3.9 - 4.2 & 94 & 17 & 271 & 49   \\
4.2 - 4.6 & 101 & 19 & 348 & 64 \\
4.6 - 4.9 & 113 & 25 & 463 & 85  \\
\hline
\end{tabular}
\end{center}
\end{table}

\section{Conclusions}
\label{sec:Conclusions}
The energy flow at large pseudorapidities, 3.15 $< |\eta| <$ 4.9, has been measured in \Pp\Pp\ collisions for minimum-bias events and events with  a dijet system in the central region, $|\eta_{jet} |< 2.5$, with $p_{T,jet}> 8$~GeV ($p_{T,jet}> 20$~GeV)
in pp collisions at $\sqrt{s} = 0.9$~TeV ($\sqrt{s} =  7$~TeV).

By requiring a high-momentum dijet in the central region, the
deposited energy increases in the entire phase space.
Thus the forward energy flow is higher in the dijet data sample than in the minimum-bias sample. The increase of the energy flow with increasing centre-of-mass energy is
reproduced in general by all the Monte Carlo event generators, both for minimum-bias and dijet events.
The results indicate that predictions of models including multiple-parton interactions are close to the data.
The Monte Carlo predictions without multiple-parton interactions,
derived from {\sc pythia}6 and {\sc cascade}, significantly underestimate the
energy flow compared to data. None of the {\sc pythia}6 tunes under study can describe all four energy-flow measurements equally well and, in general, they predict a flatter energy-flow distribution in minimum-bias events. We observe that the D6T and Pro-Q20 tunes provide the best description of the minimum-bias and dijet data, respectively. The predictions from the {\sc herwig++} tunes  as well as {\sc dipsy} are also in agreement with the data. The predictions from cosmic-ray interaction models provide the best descriptions of the measured energy flow.

The variation of the energy flow with $\eta$, both in minimum-bias and dijet events, is reasonably well reproduced by all Monte Carlo event generators with multiple-parton interactions included. However, the magnitude of the average energy strongly depends on the parameter settings of the different MC tunes, as shown by the large spread in the theoretical predictions.

A comparison with measurements at HERA in deep-inelastic \Pe\Pp\ collisions, where the contribution from multiple-parton interactions is negligible, shows that the (transverse) energy flow in \Pp\Pp\ collisions is significantly larger.

\section*{Acknowledgements}
We would like to thank Tanguy Pierog, Ralph Engel, Sergey Ostapchenko and Ralf Ulrich for providing the predictions of the cosmic-ray MC generators, Simon Pl\"{a}tzer for providing the predictions of Herwig++ and Leif L\"{o}nnblad for making the \textsc{dipsy} predictions available to the CMS collaboration.

We wish to congratulate our colleagues in the CERN accelerator departments for the excellent performance of the LHC machine.
We thank the technical and administrative staff at CERN and other CMS institutes for their devoted efforts during the design, construction and operation of CMS. The cost of the detectors, computing infrastructure, data acquisition and all other systems without which CMS would not be able to operate was supported by the financing agencies involved in the experiment.
We are particularly indebted to: the Austrian Federal Ministry
of Science and Research; the Belgium Fonds de la Recherche Scientifique, and Fonds voor Wetenschappelijk Onderzoek;
the Brazilian Funding Agencies (CNPq, CAPES, FAPERJ, and FAPESP); the Bulgarian Ministry of Education and Science; CERN;
the Chinese Academy of Sciences, Ministry of Science and Technology, and National Natural Science Foundation of China;
the Colombian Funding Agency (COLCIENCIAS); the Croatian Ministry of Science, Education and Sport;
the Research Promotion Foundation, Cyprus; the Estonian Academy of Sciences and NICPB;
the Academy of Finland, Finnish Ministry of Education, and Helsinki Institute of Physics;
the Institut National de Physique Nucl\'eaire et de Physique des Particules~/~CNRS, and Commissariat \`a l'\'Energie Atomique, France;
the Bundesministerium f\"ur Bildung und Forschung, Deutsche Forschungsgemeinschaft, and Helmholtz-Gemeinschaft Deutscher For-schungszentren, Germany;
the General Secretariat for Research and Technology, Greece; the National Scientific Research Foundation, and National Office for Research and Technology, Hungary;
 the Department of Atomic Energy, and Department of Science and Technology, India; the Institute for Studies in Theoretical Physics and Mathematics, Iran; the Science Foundation, Ireland;
the Istituto Nazionale di Fisica Nucleare, Italy; the Korean Ministry of Education, Science and Technology and the World Class University program of NRF, Korea;
the Lithuanian Academy of Sciences; the Mexican Funding Agencies (CINVESTAV, CONACYT, SEP, and UASLP-FAI); the Pakistan Atomic Energy Commission;
the State Commission for Scientific Research, Poland; the Funda\c{c}\~ao para a Ci\^encia e a Tecnologia, Portugal; JINR (Armenia, Belarus, Georgia, Ukraine, Uzbekistan);
the Ministry of Science and Technologies of the Russian Federation, and Russian Ministry of Atomic Energy;
the Ministry of Science and Technological Development of Serbia; the Ministerio de Ciencia e Innovacion, and Programa Consolider-Ingenio 2010, Spain;
the Swiss Funding Agencies (ETH Board, ETH Zurich, PSI, SNF, UniZH, Canton Zurich, and SER);
the National Science Council, Taipei; the Scientific and Technical Research Council of Turkey, and Turkish Atomic Energy Authority;
the Science and Technology Facilities Council, UK; the US Department of Energy, and the US National Science Foundation.
Individuals have received support from the Marie-Curie IEF program (European Union); the Leventis Foundation; the A. P. Sloan Foundation;
the Alexander von Humboldt Foundation; and the Associazione per lo Sviluppo Scientifico e Tecnologico del Piemonte (Italy).

\bibliography{auto_generated}   
\cleardoublepage \appendix\section{The CMS Collaboration \label{app:collab}}\begin{sloppypar}\hyphenpenalty=5000\widowpenalty=500\clubpenalty=5000\textbf{Yerevan Physics Institute,  Yerevan,  Armenia}\\*[0pt]
S.~Chatrchyan, V.~Khachatryan, A.M.~Sirunyan, A.~Tumasyan
\vskip\cmsinstskip
\textbf{Institut f\"{u}r Hochenergiephysik der OeAW,  Wien,  Austria}\\*[0pt]
W.~Adam, T.~Bergauer, M.~Dragicevic, J.~Er\"{o}, C.~Fabjan, M.~Friedl, R.~Fr\"{u}hwirth, V.M.~Ghete, J.~Hammer\cmsAuthorMark{1}, S.~H\"{a}nsel, M.~Hoch, N.~H\"{o}rmann, J.~Hrubec, M.~Jeitler, W.~Kiesenhofer, M.~Krammer, D.~Liko, I.~Mikulec, M.~Pernicka, B.~Rahbaran, H.~Rohringer, R.~Sch\"{o}fbeck, J.~Strauss, A.~Taurok, F.~Teischinger, P.~Wagner, W.~Waltenberger, G.~Walzel, E.~Widl, C.-E.~Wulz
% ...
\vskip\cmsinstskip
% ...
1:~~Also at CERN, European Organization for Nuclear Research, Geneva, Switzerland\\
% ...

\end{sloppypar}
\end{document}